Cite as:

**IEEE**
M. Jafarzadeh and Y. Tadesse, "Convolutional Neural Networks for Speech Controlled Prosthetic Hands", in *2019 First International Conference on Transdisciplinary AI (TransAI)*, Laguna Hills, California, USA, 2019, pp. 35-42.

**ACM**
Jafarzadeh, M. and Tadesse, Y., 2019. Convolutional Neural Networks for Speech Controlled Prosthetic Hands. In *2019 First International Conference on Transdisciplinary AI (TransAI)*. IEEE, pp. 35-42.

**MLA**
Jafarzadeh, Mohsen, and Yonas Tadesse. "Convolutional Neural Networks For Speech Controlled Prosthetic Hands". IEEE, *2019 First International Conference On Transdisciplinary AI (Transai)*, 2019, pp. 35-42, Accessed 26 Sept 2019.

**APA**
Jafarzadeh, M., & Tadesse, Y. (2019). Convolutional Neural Networks for Speech Controlled Prosthetic Hands. In *2019 First International Conference on Transdisciplinary AI (TransAI)* (pp. 35-42). Laguna Hills, California, USA: IEEE.

**Harvard**
Jafarzadeh, M. and Tadesse, Y. (2019). Convolutional Neural Networks for Speech Controlled Prosthetic Hands. In: *2019 First International Conference on Transdisciplinary AI (TransAI)*. IEEE, pp.35-42.

**Vancouver**
Jafarzadeh M, Tadesse Y. Convolutional Neural Networks for Speech Controlled Prosthetic Hands. 2019 First International Conference on Transdisciplinary AI (TransAI). IEEE; 2019. p. 35-42.

**Chicago**
Jafarzadeh, Mohsen, and Yonas Tadesse. 2019. "Convolutional Neural Networks For Speech Controlled Prosthetic Hands". In *2019 First International Conference On Transdisciplinary AI (Transai)*, 35-42. IEEE.

**BibTeX**
@inbook{jafarzadeh_tadesse_2019,[break] title={Convolutional Neural Networks for Speech Controlled Prosthetic Hands},[break] booktitle={2019 First International Conference on Transdisciplinary AI (TransAI)},[break] publisher={IEEE},[break] author={Jafarzadeh, Mohsen and Tadesse, Yonas},[break] year={2019},[break] pages={35-42}

# Convolutional Neural Networks for Speech Controlled Prosthetic Hands

Mohsen Jafarzadeh
Department of Electrical and Computer Engineering
The University of Texas at Dallas
Richardson, TX, USA
Mohsen.Jafarzadeh@utdallas.edu

Yonas Tadesse
Department of Mechanical Engineering
The University of Texas at Dallas
Richardson, TX, USA
Yonas.Tadesse@utdallas.edu

*Abstract*—**Speech recognition is one of the key topics in artificial intelligence, as it is one of the most common forms of communication in humans. Researchers have developed many speech-controlled prosthetic hands in the past decades, utilizing conventional speech recognition systems that use a combination of neural network and hidden Markov model. Recent advancements in general-purpose graphics processing units (GPGPUs) enable intelligent devices to run deep neural networks in real-time. Thus, state-of-the-art speech recognition systems have rapidly shifted from the paradigm of composite subsystems optimization to the paradigm of end-to-end optimization. However, a low-power embedded GPGPU cannot run these speech recognition systems in real-time. In this paper, we show the development of deep convolutional neural networks (CNN) for speech control of prosthetic hands that run in real-time on a NVIDIA Jetson TX2 developer kit. First, the device captures and converts speech into 2D features (like spectrogram). The CNN receives the 2D features and classifies the hand gestures. Finally, the hand gesture classes are sent to the prosthetic hand motion control system. The whole system is written in Python with Keras, a deep learning library that has a TensorFlow backend. Our experiments on the CNN demonstrate the 91% accuracy and 2ms running time of hand gestures (text output) from speech commands, which can be used to control the prosthetic hands in real-time.**

***Keywords-convolutional neural networks; deep learning; prosthetic hands; speech control; speech recognition***

## I. INTRODUCTION

The human hand is a very complex system that can perform many and useful tasks. The idea of developing a human-like artificial hand has a long history. Over the last decades, researchers and engineers have taken advantage of the latest technological advances to develop more dexterous, realistic, and novel hands. However, we are still far from human-like hands. Without any doubt, we can say that the development and control of a prosthetic hand is still an open problem in robotics and bioengineering. Developing a prosthetic hand requires great effort and multidisciplinary knowledge including but not limited to biology, neuroscience, mechanisms, sensors, actuators, control, and cognitive science. In this paper, we discuss only the control aspects of prosthetic hands. We refer readers to review papers for extensive discussion on prosthetic hands [1-5]. In practice, any prosthetic hand should get commands from a user in some form. Researchers have used push-button, keyboard, text, speech, electromyography, vision, or a combination of these to control prosthetic hands in the past. Humans find it is easier to communicate and express their idea via speech. Therefore, controlling devices by speech is one of the most popular ideas, especially for devices designed for disabled people. For the same reason, in this paper, we only focus on controlling prosthetic hands via speech. Fig. 1 shows the functional block diagram of the system discussed in this paper.

An important section of artificial intelligence is transforming raw speech to text, which is known as automatic speech recognition (ASR). One of the major challenges in any automatic speech recognition is huge variations in speech signals. For instance, the speech signals will be different in all repetitions with the same recording device in the same environment when the same person speaks the same phrase. If the same phrase is uttered by another person with different tones and accents, the difference will increase. Regardless of these challenges, scientists attempted to develop ASR systems that work in real-time and robust to different speakers and environments.

Traditional ASR systems consist of four subsystems: pre-processing, feature extraction, language model, and classifier [6]. Developers usually use a combination of starting and ending point detection, normalization, noise reduction, and pre-emphasizing for the pre-processing of subsystem. Then, developers define several features such that these features map the pre-processed signal into classes. These features must be robust to noise while discriminating between classes. The feature extraction subsystem computes the pre-defined features from the pre-processed signal. The performance of a traditional ASR systems is highly dependent on the feature extraction subsystem [7]. Many useful features have been proposed [8,9]. The classifier subsystem maps the features to text. For classifier, researchers have widely studied the hidden Markov model (HMM), Gaussian mixture model (GMM), multi-layer perceptron (MLP), self-organizing map (SOM), radial basis function (RBF), fuzzy neural networks (FNN), adaptive neuro-fuzzy inference system (ANFIS), recurrent neural networks (RNN), long short-term memory (LSTM), gated recurrent units (GRU), and support vector machine (SVM). Many researchers reported that a combination of artificial neural networks and hidden Markov model (ANN-HMM) is more accurate than a combination of the Gaussian mixture model and the hidden Markov model (GMM-HMM) in the case of traditional ASR systems [8,9]. In the best works in traditional ASR systems, language models are defined in the hidden Markov model and are solved by the Viterbi search algorithm. Research in ASR systems has yielded several open-source libraries, including CMU Sphinx [10], Kaldi [11], ESPnet [12], Openseq2Seq [13], and Eesen [14]. Although the performance of traditional automatic speech recognition systems increased over time, still these systems often perform poorly in noisy environments. Thus, scientists proposed many techniques to reduce noise effects. Indeed, these techniques enhance the overall performance of the systems. However, there remains a huge gap between the error in practical environments (such as outdoors or an office) and a high signal to noise ratio (SNR) environments (studio or lab condition) [15]. In brief, a critical issue of traditional automatic speech recognition systems in real scenarios is robustness.

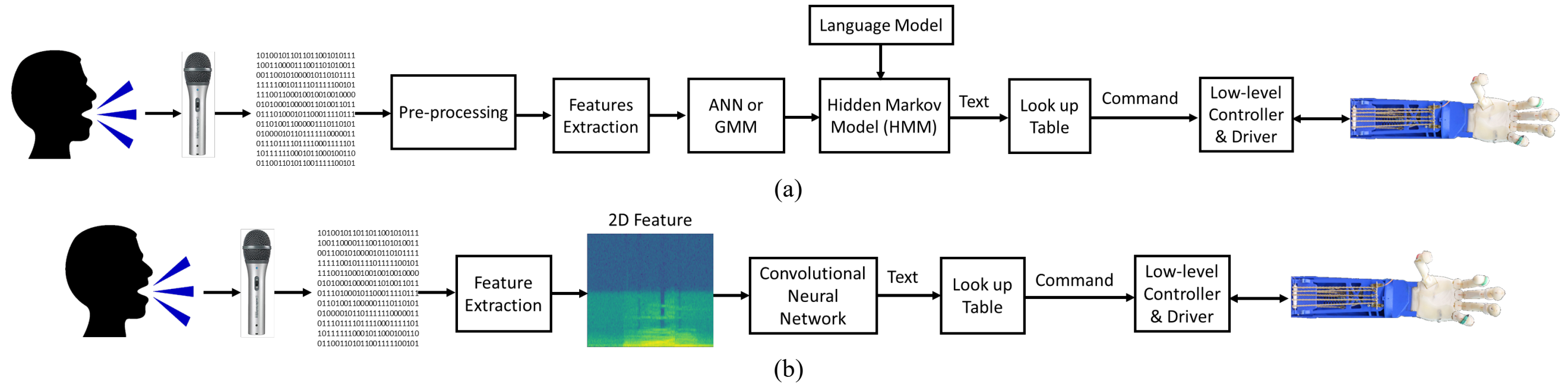

Figure 1. Block diagram of controlling prosthetic hands via speech using (a) traditional automatic speech recognition systems (b) proposed method.

Recently mobile devices can run in real-time a large set of computation in parallel by using new technologies in artificial intelligence application-specific integrated circuits (AI-ASIC) [16], general-purpose graphics processing units (GPGPU) [17], and tensor processing units (TPU) [18]. Also, a huge amount of labeled speech data has been released in the past few years. Thus, researchers can train very deep neural networks and use them in real-time scenarios, by using these recent parallel computing processors and big data. As a result, state-of-the-art automatic speech recognition systems have changed from a combination of artificial neural networks and hidden Markov model to very deep end-to-end neural networks [19-21]. The latest published researches in the end-to-end systems provide faster, higher accuracy, and lower size than traditional automatic speech recognition systems. Regardless of these advantages, the best end-to-end systems are still too large and computationally expensive for portable devices such as prosthetic hands, whose processors have limited memory and computation capabilities. Thus, to use end-to-end automatic speech recognition systems in prosthetics device additional studies are required. There exist several methods for creating smaller networks from large networks [22-24]. Most of these methods prune Artificial Neural Networks (ANN) by changing some weight of the network to zero to create a sparse network. Some of these methods prune neurons or even layers. Although these methods are very effective to reduce the size of the ANN, they are not sufficient by themselves for many mobile applications. Therefore, it is better to first find a smaller ANN then use these methods to reduce the size further.

Traditional ASR systems have been used in many projects to command prosthetic limb and other smart devices. In [25], a combination of Mel-frequency cepstral coefficients feature extractor, hidden Markov model classifier, and Viterbi has been used to control a smart home via speech. In another work, long short-term memory has been used in a voice command module [26]. To command a surgical robot by speech, researchers used CMU Sphinx [27]. To control a hand exoskeleton, developers used a combination of discrete wavelet transforms and hidden Markov models [28]. Developers used a multi-layer perceptron to control a robotic hand with 13 speech features that compromise five features in the time domain and eight features in the frequency domain [29]. To reduce the influence of external noise, they performed training and testing in an identical room.

In this paper, we focus on the control of prosthetic hands with speech input using convolutional neural network (CNN). In contrast to prior studies that use the traditional automatic speech recognition systems, we utilize only a convolutional neural network that converts 2D features (like spectrogram) of speech input to text. Thus, this is a unique approach for controlling prosthetic hands. We did not use the hidden Markov model. Many prosthetic hands have only a low-end embedded GPGPU. Therefore, the computational powers of these prosthetic hands are limited. Thus, the state-of-the-art automatic speech recognition systems that use very deep convolutional neural networks cannot run in real-time in this category of prosthetic hands because of their limited computing resources. Therefore, we tried to minimize the size of the convolutional neural network such that they can run in real-time in the prosthetic hands' processing units.

The paper is organized as follows. The proposed convolutional neural network is explained in section II. The experimental result of the proposed CNN on an embedded GPGPU is presented in section III. Low-level controllers and drivers for prosthetic hands are described in section IV. In the next section, we discuss the final products. The last section is a conclusion on the proposed convolutional neural networks for controlling prosthetic hands with embedded GPU via speech.

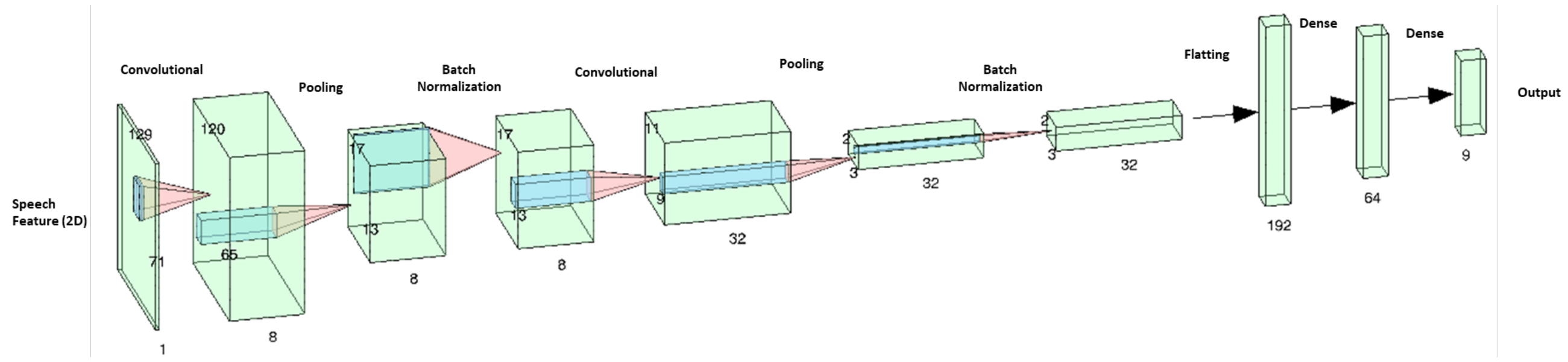

Figure 2. Block diagram of the proposed convolutional neural network (details stated in table II).

TABLE I. SUMAMARY OF THE PROPOSED CONVOLUTIONAL NEURAL NETWORKS

| Layer | Type | Number of filters | Filter size | Stride | Activation | Output shape | Number of Parameters |
|---|---|---|---|---|---|---|---|
| 0 | Input (Log of spectrogram) | - | - | - | - | 129 x 71 x 1 | 0 |
| 1 | Convolution 2D | 8 | 10 x 7 | 1 | ReLU | 120 x 65 x 8 | 568 |
| 2 | Pooling 2D | - | 7 x 5 | 1 | Max | 17 x 13 x 8 | 0 |
| 3 | Batch normalization | - | - | - | | 17 x 13 x 8 | 32 |
| 4 | Convolution 2D | 32 | 7 x 5 | 1 | ReLU | 11 x 9 x 32 | 8992 |
| 5 | Pooling 2D | - | 5 x 3 | 1 | Max | 2 x 3 x 32 | 0 |
| 6 | Batch normalization | - | - | - | - | 2 x 3 x 32 | 128 |
| 7 | Flatten | - | - | - | - | 192 | 0 |
| 8 | Dense | - | - | - | ReLU | 64 | 12352 |
| 9 | Drop out | - | - | - | - | 64 | 0 |
| 10 | Dense | - | - | - | SoftMax | 9 | 585 |

## II. METHOD

In this section, we explain the proposed lightweight convolutional neural network to control a prosthetic hand by speech command. Fig. 1 (b) demonstrate the overall system. First, a speech signal is captured by a digital microphone. Here, we assumed that the analog to digital converter (ADC) inside the digital microphone converts analog speech data to an integer (16-bit or higher). The proposed method also works with a combination of an analog microphone and an external ADC (audio adapter). The sampling frequency should be equal to or greater than 16 kHz. In the next step, a 2D features (for example spectrogram) is computed from the stream of integer numbers. The 2D features that the proposed method can use must have local correlations in both time and frequency. For example, speech spectrogram satisfies this condition (has local correlations in both time and frequency). CNNs can handle local correlation explicitly by local connectivity. Thus, it is easier to extract useful features and training system with CNNs than dense neural networks or fully connected recurrent neural networks. Therefore, we used a CNN to map the 2D features to the equivalent text. In the next step, the predicted text will be mapped to command (trajectory) by using a look-up table. For example, if the output of the proposed convolutional neural network is the word “two”, the look-up table assigns 0s (full relaxation) to the index and middle fingers and that 1s (full contraction) to the rest of the fingers. Finally, the command (trajectory for each finger) will be sent to the low-level controller and driver of the prosthetic hand.

To communicate with standard prosthetic hand drivers, we should use UART, SPI, or I2C buses (depends on the drivers). In the current market (2019), there are few GPGPU and TPU developer kits that provide these buses. Table II shows a comparison of the three most powerful GPGPU/TPU developer kits on the market. Currently, the NVIDIA AGX Xavier developer kit is the best kit. The NVIDIA Jetson TX2 [30] has a lower cost than NVIDIA AGX Xavier. The NVIDIA Jetson TX2 provides sufficient computation power. Therefore, we used NVIDIA Jetson TX2 in this paper. As a result, the proposed CNN has been optimized for NVIDIA Jetson TX2. Indeed, users can run the proposed convolutional neural network without any change in the NVIDIA AGX Xavier developer kit.

The proposed CNN maps a 2D feature (spectrogram) of the speech signal to the corresponding class (one-hot vector). The size of the network output (vector) is the number of words in the dictionary plus one for the unknown. For example, if a dictionary contains the set of {zero, one, …, five, on, off}, the dimension of the output vector will be 9. We started from a two-layer network (a convolutional layer and one dense layer), and add layers to find an optimum tradeoff between accuracy and speed. The proposed CNN architecture is similar but not identical to [31]. Therefore, it is a new network. A summary of the proposed CNN is shown in Fig. 2 and the detail is stated in Table I. We tried to optimize the hyper parameters such as number of layers, filter size, number of filters and stride. More optimization can be done in the future. The proposed network has two 2D convolution layers and two dense layers. We used max-pooling and batch normalization after each convolutional layer. The proposed network has 22577 parameters. We used a dropout layer between the two dense layers to improve the learning rate. We used rectified linear units (ReLU) in the activation function of convolutional layers and the first dense layer. We used SoftMax in the activation function of the last

TABLE II. TPU AND GPU DEVELOPER KITS COMPARISON

| Company | Google | NVIDIA | NVIDIA | NVIDIA |
|---|---|---|---|---|
| Model | Coral | Jetson Nano | Jetson TX2 | AGX Xavier |
| GPU | Vivante GC7000 Lite 16 core | Maxwell 128 core | Pascal 256 core | Volta 512 core |
| TPU | Google Edge | - | - | - |
| CPU | 4 core Cortex-A53 | 4 core Cortex-A57 | 4 core Cortex-A57 + 2 core Denver | 8 core Carmel |
| RAM | 1 GB | 4 GB | 8 GB | 16 GB |
| Storage | 8 GB | 16 GB | 32 GB | 32 GB |
| GFLOPS | **32** | **236** | **559** | **1300** |
| GPIO | 8 | 5 | 8 | 4 |
| USB | 1 x USB 3.0 + 1 x USB C | 4 x USB 3.0 | 1 x USB 3.0 +1 x USB 2.0 | 2 x USB C [3.1] |
| UART | 2 | 1 | 1 | 1 |
| I2C | 2 | 2 | 4 | 2 |
| SPI | 1 with 2 CS | 2 with 2 CS | 1 with 2 CS | 1 with 2 CS |
| CAN | 0 | 0 | 1 | 1 |
| I2S | 1 | 1 | 2 | 1 |
| Size (mm) | 88 x 60 x 24 | 100 x 80 x 29 | 170 x 170 x 51 | 105 x 105 x 85 |
| Weight | 227 g | 244 g | 1.5 Kg | 630g |
| Price ($) | 150 | 100 | 400 | 700 |

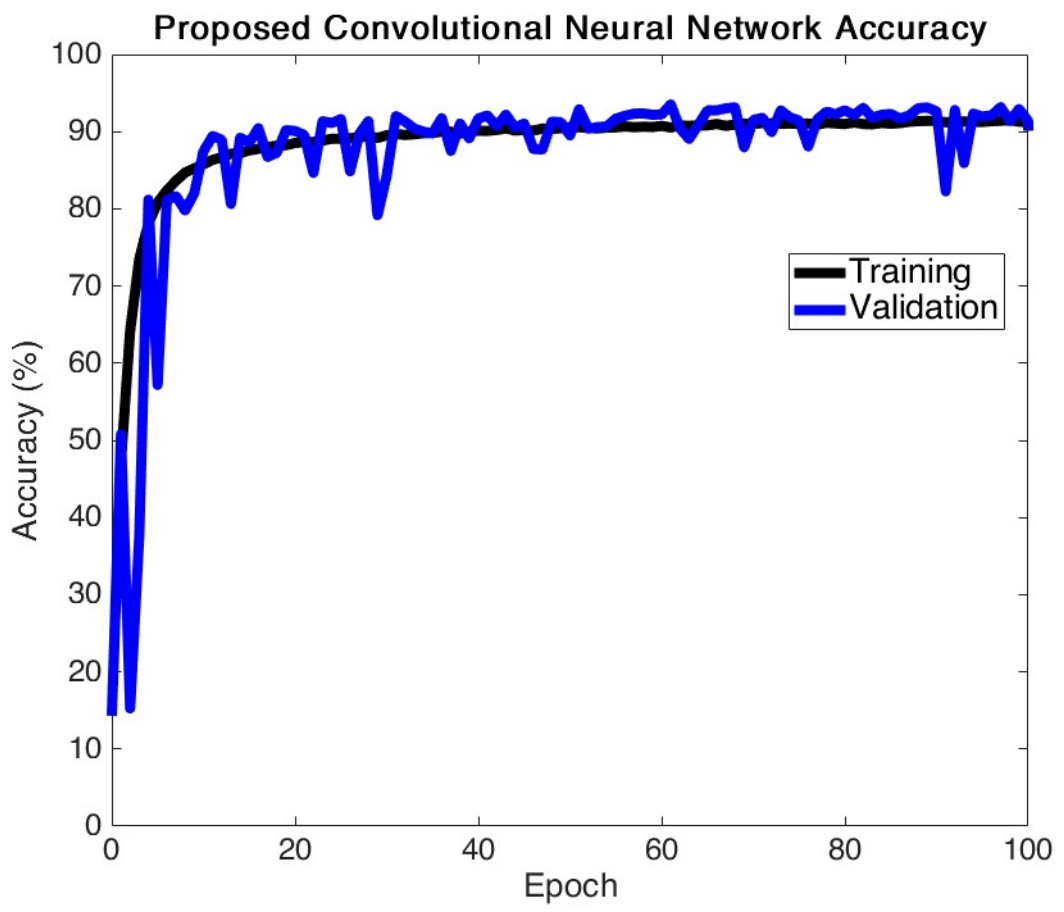

Figure 3. Training and validation accuracy with ADAM optimizer

dense layer because the output vector is the one-hot encoding form of word vectors.

## III. RESULTS

We tested our approach by creating an experimental setup. We designed a PCB (driver board) that consists of an 16-bit digital to analog converter (DAC) and power amplifiers. The custom-designed PCB communicates with a GPGPU kit via an I2C bus. The driver board applies the desired voltages that receive from the GPGPU kit to the actuators in the prosthetic hand. We used Audio-Technica ATR2100-USB (a USB digital microphone) for capturing the speech signals. Linux Ubuntu OS was running on the NVIDIA Jetson TX2 developer kit. We only used Python programming language to run the experimental setup. In addition to Python, the kit can run C and C++, if a higher speed is required. To calculate the spectrogram, we used the SciPy library. A negligible number ($10^{-10}$) has been added to the spectrogram to avoid the singularity. We used the NumPy library to compute the logarithm of the spectrogram. The proposed network receives the logarithm of the spectrogram as its input. To implement the proposed convolutional neural networks, we used the Keras library with the backend of TensorFlow.

In training ASR systems, speech data set is essential. CNNs require a huge amount of data for training and validation. We used the Google speech command data set [32]. The data set is open-source and it is released under the Creative Commons BY 4.0 license. The dataset contains a total of 35 words (105,829 utterances). Each utterance is one-second or shorter. The sample data stored as WAV format files. The data is encoded in 16 kHz rate with linear 16-bit single-channel PCM values. In addition to the 35 words, Google added several minutes long various kinds of background noise. These background noises were captured directly from noisy environments. We used these environment noise in the training to make the system robust to environmental noise. Here, we used the 8 classes (words) from 35 words, which are "zero", "one", "two", "three", "four", and "five", "on" and "off". Also, we used the rest of the words as a class, "unknown".

The NVIDIA AGX Xavier has enough RAM to train and test the proposed network. However, the NVIDIA Jetson TX2 has only 8 GB RAM. About 2 GB of RAM are used by the operating systems and other software. The remaining 6 GB on NVIDIA Jetson TX2 is not sufficient for training steps. However, it is sufficient to run and test the proposed networks. There are two solutions to use NVIDIA Jetson TX2, (1) training on external GPU such as NVIDIA Titan V (desktop GPU) and NVIDIA Tesla V100 (supercomputer GPU) and transfer the trained weights to the embedded GPGPU kit and (2) creating a 4 GB swap file. Here, we select the second option to reduce the total cost of the project. However, we suggest using the first solution when the user already has an external GPU. In the training of the proposed CNN, we used the ADAM optimizer. The accuracy of the proposed CNN was higher when we used the logarithm of the spectrogram than when we used the spectrogram directly. In this configuration, each epoch took about 93 seconds. The trend of training and validation accuracies has been illustrated in Fig. 3. The validation accuracy reached 91% in 84 epochs. Then, we tested the trained networks with more than 50 people in a crowded environment, an expo with more than 400 people in 1400 $m^2$ (15,000 $ft^2$) as preliminary evaluation. People spoke few words to the microphone and the CNN determined the words accurately in most cases (words such as on, off, three, etc). Among all the trained words, the word “four” has the most misclassification (error) during the test. We will determine the final error (average and standard deviation) for this real-world evaluation in the future work because such experiments require human subject and must be approved by Institutional Review Board (IRB). The proposed CNN runs in real-time (2 ms) on the NVIDIA Jetson TX2. We predict that it can run in 1 ms in a NVIDIA AGX Xavier. This running time is excellent to use it in prosthetic hands.

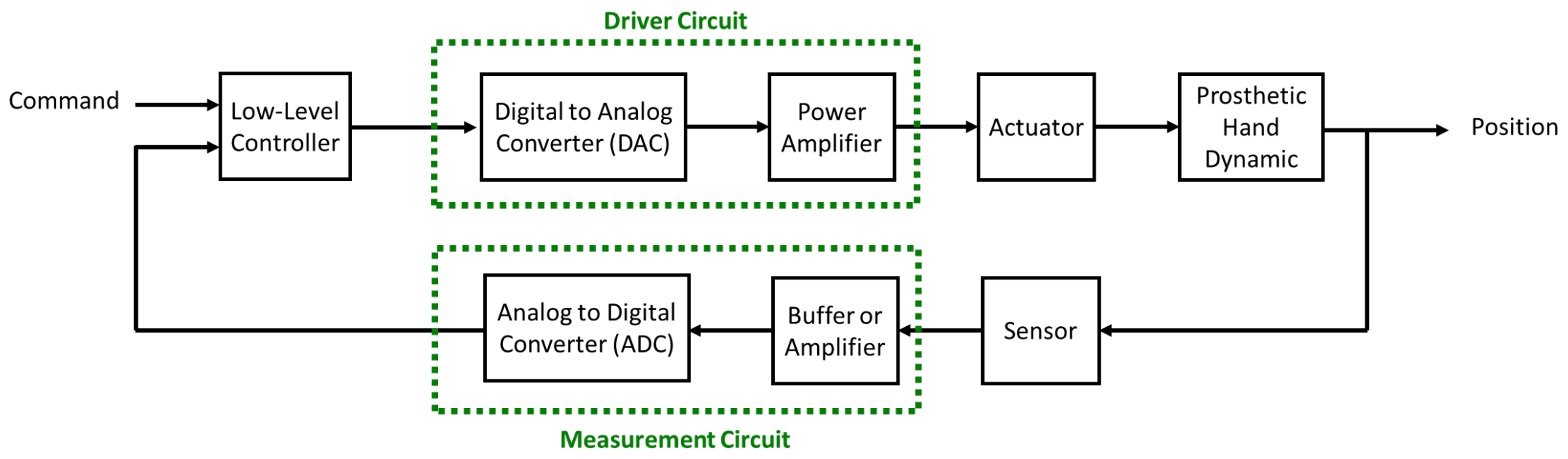

Figure 4. Block diagram of a general low-level controller and driver.

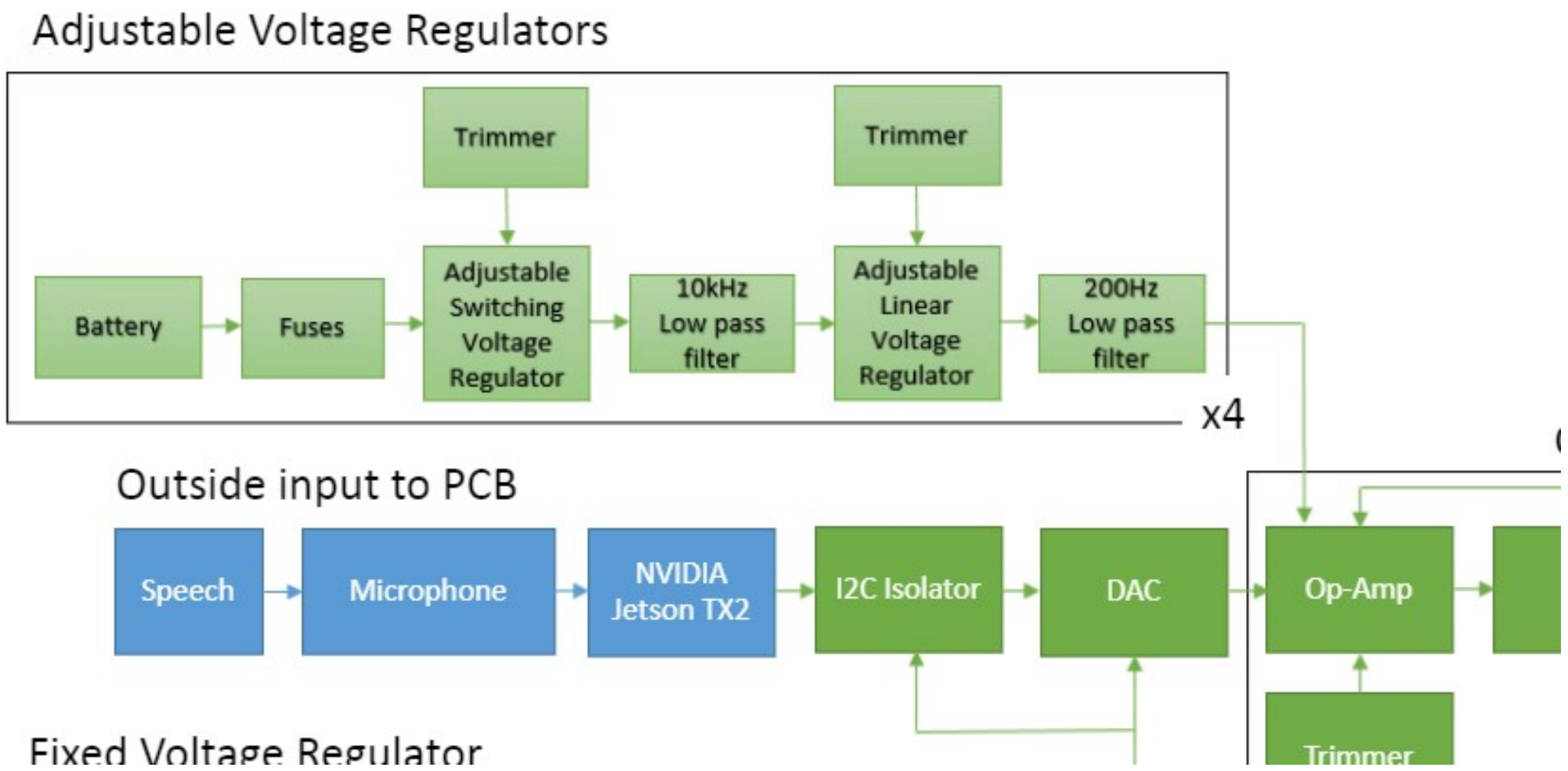

Figure 5. Block diagram of the first prototype driver circuit with four channel SMA/TCP actuators.

## IV. LOW-LEVEL CONTROLLER AND DRIVER

The proposed convolutional neural network maps the speech 2D feature to text. We used a look-up table for mapping from the text space to trajectory space (commands). As shown in Fig. 4, to guide the hand to desire position (trajectory), a low-level controller is designed. The low-level controller reads positions of the fingers and applies control law (voltage) by the driver circuit. There are many options for low-level controllers such as PID controller, Fuzzy controller, model-predictive controller, etc.

Actuators of prosthetic hands can be divided into two groups: bidirectional and unidirectional. Unidirectional actuators apply force/torque in one direction when considering electrical input voltage as a stimulus. Shape memory alloy (SMA) and twisted and coiled polymer (TCP) muscle are examples of unidirectional actuators. In the past few years, researchers at the HBS lab at the University of Texas at Dallas developed several hands that use SMA/TCP actuators. Six-ply TCPs generate enough force to be considered for prosthetic hands. Six-ply TCP actuators require a high amount of current (up to 3A), which led us to design our circuit to handle such magnitude of current. The voltage of these actuators depends on the length of the actuators and ranges from 1V up to 24 V. We used standard I2C communication protocol for communication between the GPGPU kit and the driver circuit. Fig. 5 shows the block diagram of the proposed driver circuit for four actuators and we developed our first prototype accordingly. The first prototype has only four channels and can be extended to 8 channels by repetition. The first prototype consists of 5 adjustable switching and 5 linear voltage regulators, paired low pass filters, a 16-bit digital to analog converter (DAC), 4 rail-to-rail single supply operational amplifiers (Op-Amp) for amplifying voltage, and 4 power bipolar junction transistors (BJT) for amplifying current. Fig. 6 illustrates the schematic diagram of the power amplifier (utilizing both Op-Amp and BJT) where v+ is the output of the linear regulator and respective filter. Fig. 7 shows the first prototype of the PCB. The PCB was fabricated in four layers, with green color masks, 2.0 mm thickness, 2 oz. copper weight, and an ENIG-RoHS surface finish. Table III outlines the parts in detail. The prototype circuit has five power inputs to cover the four actuator channels as well as power used by the digital components. Voltage input (a battery) of each channel and the DAC channel must be greater than 8V and less than 16V. The power efficiency of the prototype with a typical load (3.8 Ω) is 69%. The LC filters used in the assembled prototypes have a total series resistance of 1.09 Ω. The first prototype demonstrates 7.2 ms settling time. Each channel of the proposed circuit can be adjusted (by hardware without losing resolution) from 3.3 V to 15.8 V. For actuators with maximum voltage less than 3.3 V, users can set the hardware voltage to 3.3 V and apply lower voltage by software. In these cases, the resolutions drop proportionally, for example, 0.825 V actuator can be used with 14-bits of resolution (nominally, i.e. without considering noise). The resolution of the prototype circuit is 8-bit because of the limitations of the filters. The output resolutions will increase to 16-bits if we replace (1) MIC29503WT (the linear voltage regulator) with another linear voltage regulator that has lower output noise such as Texas Instruments LM350A and (2) the 200 Hz low-pass filter with 20 Hz low-pass filter.

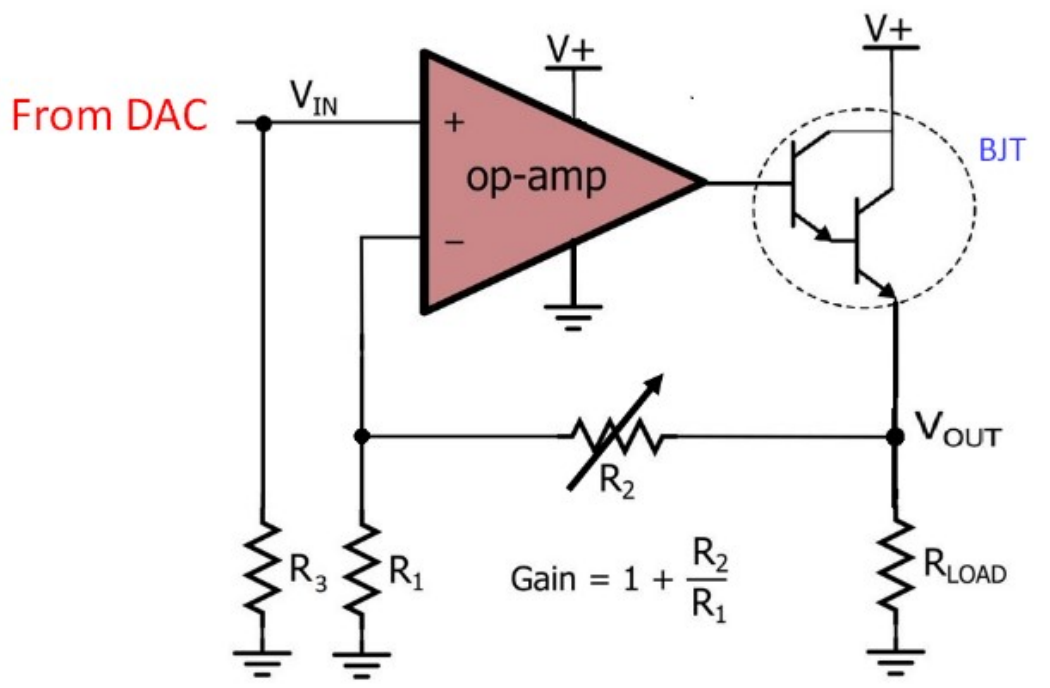

Figure 6. Schematic diagram of a power amplifier with rail-to-rail single supply Op Amp and Power NPN bipolar junction transistor.

There are many advantages of our circuit. The first one is that each channel has individual power regulation circuits, allowing them to be used individually if needed for a different robotic application. The second one is the nature of the signals allows the board to be used for any unidirectional actuators, in addition to different configurations of the TCP actuators used in our experiments. The first prototype has three main disadvantages, which are the size, the limited number of channels, and the output resolutions. The size of the first prototype circuit is large (7 x 7 inches) because it was designed to work with wide range voltages. If we design for a specific voltage (for a hand, not all the hands), we can drop adjustable switching regulators; then, it will reduce the PCB size by half. The second reason is that we used through-hole components, which are normal for the first prototype. Our next goals in the future are to design a second prototype (1) without switching regulators, (2) with 8 channels, (3) 16-bit output resolutions and (4) using surface mount components. In the next prototype, we will use a NVIDIA AGX Xavier developer kit instead of the NVIDIA Jetson TX2 developer kit. We will design the measurement circuit for a set of position sensors after fabricating a second prototype.

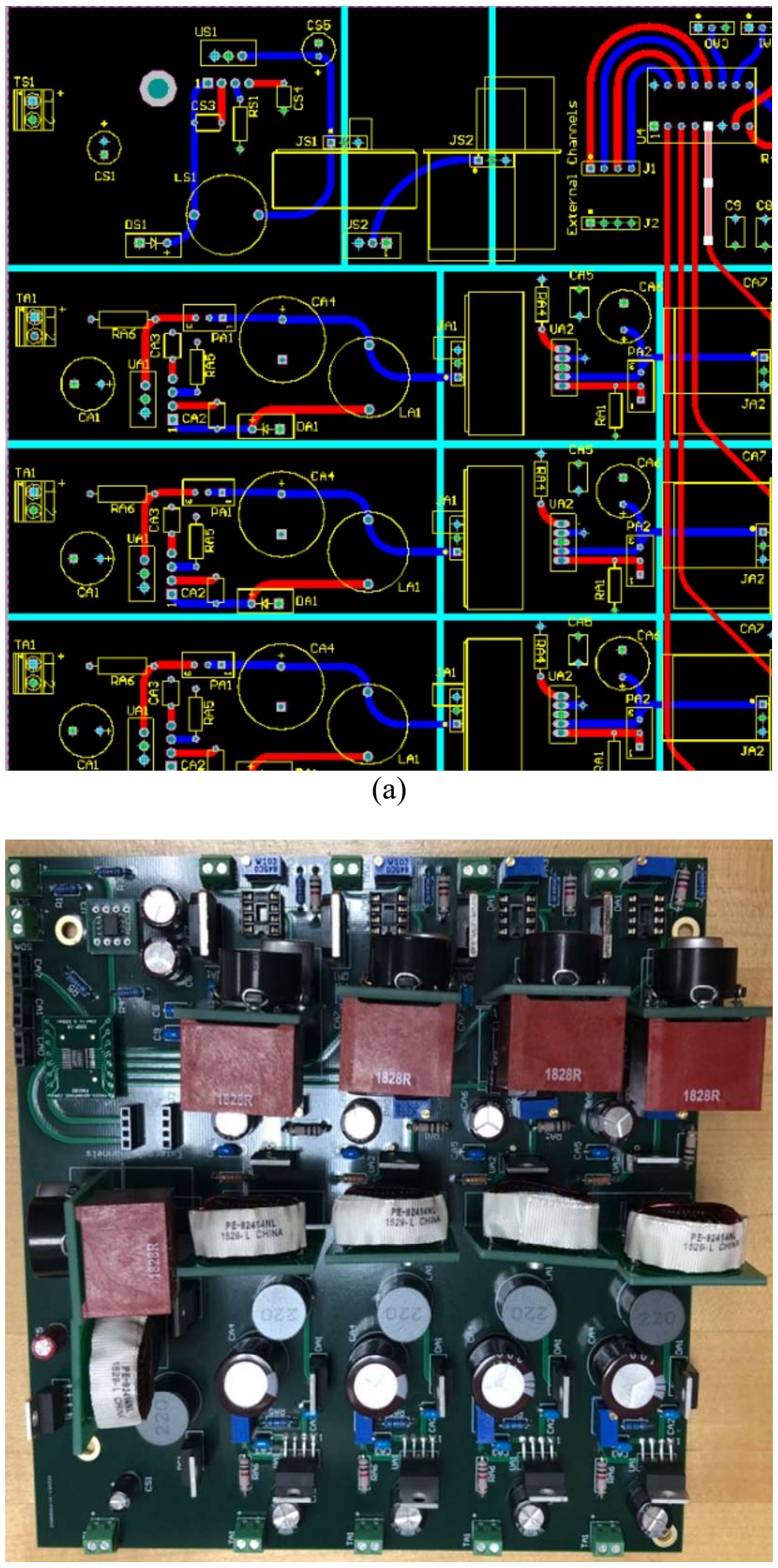

(a)

(b)

Figure 7. The first prototype of the proposed driver circuit for SMA/TCP actuators (a) CAD model (b) assembled prototype.

TABLE III. SUMAMARY OF THE COMPONENTS IN THE FIRST PROTOTYPE OF THE DRIVER CIRCUITS

| Components | Model | Description |
|---|---|---|
| DAC | LTC2605IGN-1#PBF | 16-bit, I2C, 8 channels |
| I2C Isolator | TI ISO1540QDQ1 | 1Mbps 25kV/µs |
| Op Amp | TI OPA2241 | Single-Supply, rail-to-rail output (50 mV) |
| BJT | Sanken 2SD2560 | NPN, DC Current Gain 5000, Power Max 130W |
| Adjustable switching voltage regulator | TI LM2679T-ADJ/NOPB | 5A, 60-kHz, Min input 8V |
| Adjustable linear voltage regulator | MIC29503WT | Output 5A, dropout 0.6V |
| 5.0 V switching voltage regulator | TI LM2679T-5.0/NOPB | 5A, 60-kHz, Min input 8V |
| 3.3 V linear voltage regulator | TI LM1085IT-3.3/NOPB | load regulation 0.1%, 3A |

## V. Discussions and future works

In this paper, we proposed a CNN that maps the speech signal to 9 classes (8 words plus unknown). It uses the one-hot encoding. Therefore, users can increase the number of classes by increasing the dimension output vector. Although our experiments show that increasing the depth of the network has little effect on overall accuracy, the user may need to increase the number of filters, if the number of words increases significantly.

We consider a speech recognition system is real-time, if it runs in 10 ms or less. We used Python and Keras in this work. In our experiment on NVIDIA Jetson TX2 (mid-level GPU), the running time is about 2 ms. The running time can further decease by using a better GPGPU (NVIDIA AGX Xavier). The speed of the system will be slightly decreased by adding more words. Using C++ and TensorRT [34] will increase the speed of the proposed method. However, TensorRT will decrease the accuracy of the system a little. In the future, we will study the effect of TensorRT. WAV2LETTER++ is an open-source C++ API for the end-to-end speech recognition system, which has been released recently [33]. They claim that their API is faster than other existing open-source APIs. However, they did not test their API on an embedded GPGPU. We plan to test the speed of the proposed CNN using WAV2LETTER++.

Our proposed CNN is robust to accent, speed, noise, etc. Therefore, users with diverse voices can control the prosthetic hand. However, this is not desired because we only want the owner of the prosthetic hand to control it. To address this limitation, in the future works, we plan to investigate a CNN that detects the owner's voice from other speakers. Then by combining these two CNNs, the prosthetic hand will respond only to the owner.

As mentioned in previous sections, one-hot encoding was used in the proposed CNN. In the one-hot encoding, the desired output vector elements are zeros except an element, which is one. In other words, the element corresponding to the correct classification should have a value of one, and all other classes should have a value of zero. The one-hot encoding is crisp, i.e it is not completely matched with the reality. In fact, labels are not purely zero or one. For example, the words "one" and "on" have more correlation than the word "one" and "zero". The Teacher-Student (T/S) training approach is an encoding technique used in the past few years [35]. In the Teacher-Student approach, first, one-hot encoding is used to train a very deep neural network (teacher). Then, a smaller neural network (student) is trained using the output of the teacher network (SoftMax vector). In many cases, the student network shows more accuracy than a similar network that trains with one-hot encoding. Recently, conditional teacher-student training [36] shows more accuracy than the original (unconditional) teacher-student method. In the conditional teacher-student method, first, the student network is trained with teacher output; then, the student network is trained with one-hot encoding. We plan to test the unconditional and conditional teacher-student training methods with the proposed CNN in the future.

In this project, the logarithm of speech spectrogram has been used as the 2D feature input of the proposed CNN. When the signal to noise ratio is greater 10, the logarithm of speech spectrogram works very well. However, the speech spectrogram is sensitive to high noise. To achieve a high signal to noise ratio, a cardioid polar pattern microphone has been used. Researchers found that power-normalized cepstral coefficients (PNCC) is more robust to noise than the spectrogram [37]. Therefore, we plan to test the proposed CNN with PNCC in the future.

The proposed CNN uses 2D features of speech signal as the input. Therefore, the CNN consists of 2D convolutional and 2D max polling layers. Indeed, some information has been lost during the conversion from raw speech to the 2D features. Recently, few CNNs with raw speech input that uses 1D convolutional and 1D pooling layers, have been proposed [38-40]. We will investigate to use the raw speech signal as the input in the future work. We will use NVIDIA AGX Xavier and will try to find a lighter convolutional neural network with raw speech input to use in prosthetic hands.

## VI. Conclusion

In this paper, we demonstrated speech control of a prosthetic hand using a convolutional neural network. First, a digital microphone records the speech signal. Then, the speech signal is converted to a 2D feature. The proposed CNN predicts the probability of classes corresponding to the 2D feature. The network's output (the class with maximum likelihood) is mapped to the trajectory (command) by a look-up table. Finally, the command is sent to the hand low-level controller and driver. We used NumPy, SciPy, TensorFlow, and Keras libraries in Python to develop the software. We used the NVIDIA Jetson TX2 developer kit (an embedded GPGPU) to train and test the proposed network. In this configuration, it runs in 2 ms (real-time) with the accuracy of 91% in a noisy environment.

In the future works, we will examine the running-time and accuracy of current implementation (Keras library in Python programming language) with TensorRT and WAV2LETTER++ implementations, which are written in C++. Moreover, we will test the accuracy of the unconditional and conditional teacher-student training methods. Also, we will try to find new lightweight CNN that directly receives raw speech signals. Finally, we will add another CNN such that the combination of two networks responds only to the owner of the prosthetic hand.